\documentclass[a4paper]{article}

\usepackage{INTERSPEECH2021}

\usepackage{cite}
\usepackage{multirow}
\usepackage{algorithm,algorithmicx,algpseudocode}
\usepackage{graphics}
\usepackage{arydshln}
\usepackage{eqparbox}
\usepackage[subrefformat=parens]{subcaption}
\usepackage{setspace}
\usepackage{etoolbox,siunitx}
\robustify\bfseries

\let\OLDthebibliography\thebibliography
\renewcommand\thebibliography[1]{
  \OLDthebibliography{#1}
  \setlength{\parskip}{0pt}
  \setlength{\itemsep}{0pt plus 0.3ex}
}

\makeatletter
\def\adl@drawiv#1#2#3{%
        \hskip.5\tabcolsep
        \xleaders#3{#2.5\@tempdimb #1{1}#2.5\@tempdimb}%
                #2\z@ plus1fil minus1fil\relax
        \hskip.5\tabcolsep}
\newcommand{\cdashlinelr}[1]{%
  \noalign{\vskip\aboverulesep
           \global\let\@dashdrawstore\adl@draw
           \global\let\adl@draw\adl@drawiv}
  \cdashline{#1}
  \noalign{\global\let\adl@draw\@dashdrawstore
           \vskip\belowrulesep}}
\makeatother

\DeclareMathOperator*{\argmax}{argmax}

\title{Momentum Pseudo-Labeling for Semi-Supervised Speech Recognition}
\name{Yosuke Higuchi$^{1,2*}$\thanks{$^*$Research conducted during an internship at MERL}, Niko Moritz$^1$, Jonathan Le Roux$^1$, Takaaki Hori$^1$}
\address{
  $^1$Mitsubishi Electric Research Laboratories (MERL), USA\ \ 
  $^2$Waseda University, Japan}
\email{higuchi@pcl.cs.waseda.ac.jp, \{moritz, leroux, thori\}@merl.com}

\begin{document}

\maketitle
\setlength{\abovedisplayskip}{2pt}
\setlength{\belowdisplayskip}{2pt}
\setlength{\textfloatsep}{0.2cm} %
\begin{abstract}
Pseudo-labeling (PL) has been shown to be effective in semi-supervised automatic speech recognition (ASR), 
where a base model is self-trained with pseudo-labels generated from unlabeled data. 
While PL can be further improved by iteratively updating pseudo-labels as the model evolves, 
most of the previous approaches involve 
inefficient retraining of the model or 
intricate control of the label update. 
We present \textit{momentum pseudo-labeling} (MPL), 
a simple yet effective strategy for semi-supervised ASR. 
MPL consists of a pair of \textit{online} and \textit{offline} models that interact and learn from each other, inspired by the mean teacher method. 
The online model is trained to predict pseudo-labels generated on the fly by the offline model. 
The offline model maintains a momentum-based moving average of the online model. 
MPL is performed in a single training process and the interaction between the two models effectively helps them reinforce each other to improve the ASR performance. 
We apply MPL to an end-to-end ASR model based on the connectionist temporal classification. 
The experimental results demonstrate that MPL 
effectively improves over the base model and 
is scalable to different semi-supervised scenarios with varying amounts of data or domain mismatch.
\end{abstract}

\section{Introduction}
Advances in deep learning have led to remarkable success in automatic speech recognition (ASR)~\cite{hinton2012deep, graves2013speech}.
Much of the recent progress lies in the end-to-end (E2E) framework~\cite{graves2014towards, chorowski2015attention, chan2016listen}, 
which directly optimizes speech-to-text conversion and 
greatly simplifies model training and inference processes. 
While E2E systems compete with traditional hidden Markov model-based systems~\cite{chiu2018state, luscher2019rwth, karita2019comparative}, 
their performance heavily relies on the availability of a large amount of labeled data (speech-text pairs), 
which requires great annotation costs and is not always achievable. %

In order to compensate for the lack of labeled data, 
semi-supervised learning has been attracting increasing attention for improving E2E ASR. 
Semi-supervised learning utilizes labeled data as well as unlabeled data during model training, 
where the amount of labeled data is in general much smaller than that of  unlabeled data. 
Some early works for semi-supervised E2E ASR focus on training with a reconstruction framework, 
including approaches based on 
a text-to-speech model\cite{tjandra2017listening, hori2019cycle, baskar2019semi} or
a sequential auto-encoder~\cite{karita2018semi, drexler2018combining, ren2019almost}. 
Others adopted self-supervised pre-training techniques, such as 
BERT-like mask prediction~\cite{baevski2019effectiveness, ling2020deep}, 
contrastive loss~\cite{schneider2019wav2vec, chung2020generative}, and 
feature clustering~\cite{baevski2019vq, hsu2020hubert}, 
to promote downstream E2E ASR tasks. 

We focus on self-training~\cite{scudder1965probability} or \textit{pseudo-labeling} (PL)~\cite{lee2013pseudo},
which has recently been adopted for %
semi-supervised E2E ASR and shown to be effective~\cite{kahn2020self, masumura2020sequence, xu2020iterative, chen2020semi, park2020improved, weninger2020semi, hsu2020semi, likhomanenko2020slimipl, khurana2021unsupervised, moritz2021semi}. 
In PL, 
a teacher (base) model is first trained on labeled data and 
used to generate pseudo-labels for unlabeled data. 
A student model is then trained using both the labeled and the pseudo-labeled data 
to obtain better performance than the teacher. 
To obtain more sophisticated pseudo-labels, 
external language models (LMs) and beam-search decoding are often incorporated into the labeling process~\cite{kahn2020self, hsu2020semi}. 
Data augmentation is also useful for helping the student model with training on the pseudo-labeled data~\cite{chen2020semi, park2020improved, weninger2020semi}. 
In addition to these extensions of PL, 
iterative pseudo-labeling (IPL), 
where the student model (initialized with the teacher model) is continuously trained on iteratively updated pseudo-labels, has shown promising results~\cite{chen2020semi, xu2020iterative, likhomanenko2020slimipl, khurana2021unsupervised}. 
In~\cite{chen2020semi}, 
the student model generates pseudo-labels on the fly from unlabeled data in a self-supervised manner. 
Pseudo-labels are refined as the model learns and the model is improved using the continuously updating pseudo-labels. 
However, 
we found this framework unstable when there is a large amount of unlabeled 
data or a domain mismatch between labeled and unlabeled data, 
which is likely to be the case in real-world scenarios. 
While \cite{likhomanenko2020slimipl} successfully scales IPL to a large amount of data, some heuristic controls are required for updating pseudo-labels. 

In this paper, 
we present a simple, efficient, stable, and scalable semi-supervised learning algorithm for E2E ASR, 
referred to as \textit{momentum pseudo labeling} (MPL). 
In MPL, 
the pseudo-labels are iteratively updated based on 
an ensemble of models at different time steps within a single training process~\cite{laine2016temporal}. MPL consists of \textit{online} and \textit{offline} models that interact and learn from each other, 
similar to the teacher-student framework in the mean teacher method~\cite{tarvainen2017mean}. 
The online model is trained to predict pseudo-labels generated on the fly by the offline model. 
The offline model maintains a momentum-based moving average of the weights of the online model, 
which can be regarded as an ensemble of the online models at different training steps. 
Through the interaction between the two models, 
MPL effectively stabilizes the training with unlabeled data and handles the constant change in pseudo-labels. 
The advantages of the proposed MPL are summarized as follows:

\noindent\textbf{Simple and efficient:} 
The semi-supervised training is performed in a single stage, 
where pseudo-labels are generated on the fly and naturally refined without requiring an LM or beam-search decoding. 
Our approach is easy to implement and 
we show an effective way for controlling the momentum update 
which reduces the burden for heuristic tuning. 

\noindent\textbf{Stable and scalable:} 
We show the effectiveness of MPL in a variety of semi-supervised scenarios.
Our approach is robust to variations in the amount of data and 
domain mismatch severity, which often pose difficulties in semi-supervised ASR~\cite{khurana2021unsupervised}.

\section{Momentum Pseudo-Labeling}
The overall process of the proposed momentum pseudo-labeling is shown in Algorithm \ref{algo:mpl}. 
\begin{algorithm}[t]
    \caption{\bf Momentum Pseudo-Labeling}
    \label{algo:mpl}
    \begin{algorithmic}[1]
    \footnotesize
    \renewcommand{\algorithmicrequire}{\textbf{Input:}}
    \renewcommand{\algorithmicensure}{\textbf{Output:}}
        \Statex \textbf{Input:}
        \Statex \ \ \ $\mathcal{D}_{\mathrm{sup}}, \mathcal{D}_{\mathrm{unsup}}$ \Comment{labeled and unlabeled data}
        \Statex \ \ \ $\mathcal{A}$  \Comment{an ASR model architecture}
        \Statex \ \ \ $\alpha$ \Comment{a momentum coefficient}
        \State Train a base model $P_{\theta}$ with architecture $\mathcal{A}$ on $\mathcal{D}_{\mathrm{sup}}$ using \eqref{eq:l_sup}
        \State Initialize an online model $P_{\xi}$ and an offline model $P_{\phi}$ with $P_{\theta}$
        \Repeat
        \ForAll {$S \in \mathcal{D}_{\mathrm{sup}} \cup \mathcal{D}_{\mathrm{unsup}}$}
            \State Obtain $X \sim S$
            \State Obtain $Y = \begin{cases}
                Y \sim S & (S \in \mathcal{D}_{\mathrm{sup}})\\
                \hat{Y} \sim P_{\phi} (Y | X) & (S \in \mathcal{D}_{\mathrm{unsup}})
            \end{cases}$
            \State Compute loss $\mathcal{L}$ for $P_{\xi}(Y|X)$ with \eqref{eq:l_sup} or \eqref{eq:l_unsup}
            \State Update $\xi$ using ${\nabla}_{\xi}\mathcal{L}$
            \State Update $\phi \leftarrow \alpha \phi + (1 - \alpha) \xi$
        \EndFor
        \Until \textit{maximum iterations are reached} \\
        \Return $P_\xi, P_\phi$
    \end{algorithmic}
\end{algorithm}
We describe MPL in two steps: 
1) the supervised training process of a base E2E ASR model, and 
2) the proposed semi-supervised training scheme for improving the model using unlabeled data. 

\vspace{-.2cm}
\subsection{Supervised training of a base model}
\vspace{-.1cm}
The objective of E2E ASR is to model a sequence mapping between a $T$-length input sequence $X \!=\! (\bm{\mathrm{x}}_t \in \mathbb{R}^D| t\!=\!1,\dots,T)$ and an $L$-length output sequence $Y \!=\! ( y_l \in \mathcal{V} | l\!=\!1,\dots,L )$.
Here, $\bm{\mathrm{x}}_t$ is a $D$-dimensional acoustic feature at frame $t$, 
$y_l$ an output token at position $l$, and $\mathcal{V}$ a vocabulary. 
We focus on models based on the connectionist temporal classification (CTC)~\cite{graves2006connectionist, graves2014towards}, 
which has recently been revisited thanks to advances in neural network architectures~\cite{kriman2020quartznet, higuchi2020mask}. 
CTC predicts a frame-level sequence $Z=(z_t \in \mathcal{V} \cup \{\epsilon\}| t=1,\dots,T)$, which is obtained by introducing a special blank token $\epsilon$ into the output sequence $Y$. 
Based on a conditional independence assumption per frame, 
CTC models the conditional probability $P(Y|X)$ by marginalizing over all paths (frame alignments) as: 
\begin{equation}
    \label{eq:p_ctc}
    P (Y | X) = \sum_{Z \in \mathcal{B}^{-1} (Y)} \prod_{t=1}^{T} P (z_t | X), 
\end{equation}
where $\mathcal{B}^{-1}(Y)$ denotes all possible paths compatible with $Y$.

Given labeled data $\mathcal{D}_{\mathrm{sup}} \!=\! \{ (X_n, Y_n) | n\!=\!1,\dots,N \}$, 
a base model $P_\theta$ with parameters $\theta$ is trained with the supervised objective defined by maximum likelihood estimation of \eqref{eq:p_ctc}: 
\begin{equation}
    \label{eq:l_sup}
    \mathcal{L}_{\mathrm{sup}} (\theta) = - \log P_{\theta}(Y_n | A(X_n)), 
\end{equation}
where $A(\cdot)$ denotes a data augmentation for improving generalization of the model, here SpecAugment~\cite{park2019specaugment}. 

\vspace{-.2cm}
\subsection{Semi-supervised training with MPL}
\vspace{-.1cm}
\label{ssec:ssl}
The goal of semi-supervised training is to enhance the base model (trained on labeled data $\mathcal{D}_{\mathrm{sup}}$) by making good use of unlabeled data $\mathcal{D}_{\mathrm{unsup}} \!=\! \{ X_m | m\!=\!N\!+\!1,\dots,N\!+\!M \}$. 
MPL achieves this goal through 
an interaction between \textit{online} and \textit{offline} models. 
Let us define the online and offline models as $P_{\xi}$ and $P_{\phi}$, 
with model parameters $\xi$ and $\phi$, respectively. 
Both models are initialized with the trained base model $P_{\theta}$. 

On unlabeled data $X\!\in\!\mathcal{D}_{\mathrm{unsup}}$, the online model is trained using pseudo-labels $\hat{Y}$ generated by the offline model: 
\begin{equation}
    \label{eq:y_hat}
    \hat{Y} = \argmax_{Y} P_{\phi} (Y | X), 
\end{equation}
where $\argmax$ is performed based on the best path decoding of CTC~\cite{graves2006connectionist}. 
Specifically, the most probable tokens are selected at each frame and 
an output sequence is obtained by suppressing repeated tokens and removing blank symbols. 
To generate pseudo-labels with higher quality, 
beam-search decoding~\cite{chen2020semi} or an LM~\cite{hsu2020semi} are often incorporated into \eqref{eq:y_hat}, but 
we used neither of the techniques. 
We observed that beam-search decoding without an LM has little impact
on ASR accuracy for CTC-based models. %
While exploitation of a strong LM plays an important role for pseudo-labeling, 
we adopt the greedy decoding to keep our approach efficient and 
avoid over-fitting to LM information as reported in~\cite{khurana2021unsupervised, likhomanenko2020slimipl}. 
{\setlength{\abovedisplayskip}{4pt}
\setlength{\belowdisplayskip}{4pt}
With unlabeled data $\mathcal{D}_{\mathrm{unsup}}$ and the corresponding pseudo-labels from \eqref{eq:y_hat}, 
the objective of the online model is defined in the same manner as \eqref{eq:l_sup}: 
\begin{equation}
    \label{eq:l_unsup}
    \mathcal{L}_{\mathrm{unsup}} (\xi) = - \log P_{\xi}(\hat{Y}_{N+m} | A(X_{N+m})),
\end{equation}
where $\mathcal{L}_{\mathrm{unsup}}$ is maximized via a gradient descent optimization. 
Note that in \eqref{eq:l_unsup}, 
we apply the data augmentation to an unlabeled input as in~\cite{chen2020semi, masumura2020sequence}, 
aiming for the online model to learn robust prediction of pseudo-labels from the noisy input.
In Sec.\ \ref{ssec:exp_data_augmentation}, 
we show that data augmentation is an important factor of MPL.}

Assuming labeled data $\mathcal{D}_{\mathrm{sup}}$ is available during the semi-supervised process, 
the supervised loss $\mathcal{L}_{\mathrm{sup}}(\xi)$ can be incorporated into the training, 
helping stabilize the online model as it learns from unlabeled data with $\mathcal{L}_{\mathrm{unsup}}(\xi)$. 
In Sec.\ \ref{ssec:exp_momentum_update}, 
we demonstrate that MPL is also effective even when trained solely on unlabeled data (i.e., trained only with \eqref{eq:l_unsup}). 

The offline model, on the other hand, 
accumulates weights of the online model after every update via
\begin{equation}
    \label{eq:ema}
    \phi \leftarrow \alpha \phi + (1 - \alpha) \xi, 
\end{equation}
a momentum-based moving average with a momentum coefficient $\alpha \!\in\! [0, 1]$. 
This momentum update makes the offline model evolve more smoothly than the online model. 
We can thus control the change in pseudo-labels generated on the fly by the offline model at each training step. 
This is important to prevent pseudo-labels from deviating too quickly from the initial labels generated by the base model and to avoid collapsing to a trivial solution. 
Indeed, we empirically observe that training is prone to collapse (emitting only blank symbols for unlabeled data) for $\alpha\!=\!0.0$, %
in which case
the online and offline models share parameters and 
the online model is trained with self-generated pseudo-labels as in~\cite{chen2020semi}. 
The problem is prominent when there is a domain mismatch between labeled and unlabeled data, 
as is often the case in real-world deployment. 
We demonstrate the momentum update's importance in Sec.\ \ref{ssec:exp_momentum_update}. 

After training with MPL, both the online and offline models can be used for evaluation, although we use the online model as our default. Their performance is compared in Sec.~\ref{ssec:on_vs_off}.

\noindent\textbf{Tuning the momentum coefficient:} Instead of directly tuning $\alpha$ in \eqref{eq:ema}, 
we design a novel, more intuitive method for deriving an appropriate value of $\alpha$. 
Based on \eqref{eq:ema}, 
the parameters of the offline model after $K$ iterations can be computed as follows: 
\begin{equation}
    \label{eq:phi_K}
    \phi^{(K)} = \alpha^K \phi^{(0)} + (1-\alpha)\sum_{k=1}^{K} \alpha^{K-k} \xi^{(k)},  
\end{equation}
where $\phi^{(k)}$ and $\xi^{(k)}$ denote the parameters of each model at the $k$-th iteration, and
$\phi^{(0)} \!=\! \xi^{(0)} \!=\! \theta$. 
We here assume that it is important to retain some influence of the base model to stabilize the pseudo-label generation. 
As a measure of this influence, 
we focus on the term $\alpha^K \phi^{(0)}$ in \eqref{eq:phi_K} 
and 
define a weight $w$ of the base model in $\phi^{(K)}$ as $w \!=\! \alpha^K$, 
where we consider $K$ as the number of iterations (i.e., batches) in a training epoch. 
As $K$ can often be in the thousands, 
small changes in $\alpha$ lead to huge differences in $w$ (e.g., $0.999^{3000} \ll 0.9997^{3000}$), 
requiring small adjustments on $\alpha$ for different amounts of training data.
Instead of directly tuning $\alpha$ for the momentum update, 
we propose to tune the weight $w$, which can be regarded as the proportion of the base model retained after a training epoch. 
Given $w$ and $K$, $\alpha$ is calculated as $\alpha = e^{(1/K) \log w}$. 
By controlling the update through $w$, 
we expect MPL to be less affected by the amount of training data, 
which we examine in Sec.\ \ref{ssec:exp_momentum_update}. 

\noindent\textbf{Relationship to prior work:} Our approach can be considered as a variant of the self-ensembling technique~\cite{laine2016temporal} for semi-supervised learning, 
where a model is trained based on an ensemble of the models at different training steps. 
Particularly, MPL is inspired by and similar to the mean teacher framework~\cite{tarvainen2017mean}, 
in that model ensembling is performed via a moving average. %
However, MPL differs from prior work in the following perspectives. 
1) MPL uses hard (pseudo-)labels for training with unlabeled data: 
while soft labels generally contain richer information for promoting a model training, 
applying a distillation loss to CTC-based ASR systems is known to be problematic~\cite{takashima2018investigation}; as CTC models emit spiky posterior distributions and 
predictions are naturally high-confidence, 
we consider hard labels more suitable for MPL. 
2) MPL is a semi-supervised learning framework for E2E ASR: 
while most prior works focus on classification problems, 
few have introduced self-ensembling techniques to semi-supervised sequence-to-sequence mapping problems. %
3) MPL applies data augmentation (i.e., SpecAugment) to the input only for training the online model, while the offline model generates pseudo-labels in inference mode: 
since we do not use soft labels in MPL, 
it is preferable for pseudo-labels to be accurate; 
moreover, 
the online model can learn to robustly predict pseudo-labels from noisy input, 
an effective approach known as consistency training~\cite{xie2019unsupervised, chen2020semi, masumura2020sequence}. 

\vspace{-.1cm}
\section{Experiments}
\vspace{-.1cm}
\subsection{Experimental setup}
\vspace{-.1cm}
\noindent\textbf{Data:} The experiments were carried out using the LibriSpeech (LS)~\cite{panayotov2015librispeech} and TEDLIUM3 (TED3)~\cite{hernandez2018ted} datasets. 
LS consists of utterances from read English audio books and contains 960 hours of training data  (split into \texttt{\footnotesize train-clean-100}, \texttt{\footnotesize train-clean-360}, and \texttt{\footnotesize train-other-500}). 
TED3 consists of utterances from English Ted Talks and contains 450 hours of training data (\texttt{\footnotesize train-ted3}). 
We used the standard validation and test sets for each dataset. 
As input speech features, 
we extracted 80 mel-scale filterbank coefficients with three-dimensional pitch features using Kaldi~\cite{povey2011kaldi}. 
We used 1k word-pieces for tokenizing output texts, 
which were constructed from the \texttt{\footnotesize train-clean-100} texts using SentencePiece~\cite{kudo2018subword}.

\noindent\textbf{Semi-supervised settings:} 
We used \texttt{\footnotesize train-clean-100} for supervised training and 
treated the other training sets as unlabeled. 
Based on the base model trained on \texttt{\footnotesize train-clean-100} (LS-100), 
we considered different semi-supervised settings: 
LS-100/LS-360, 
an in-domain setting with unlabeled \texttt{\footnotesize train-clean-360}; 
LS-100/LS-860, 
an in-domain setting with unlabeled \texttt{\footnotesize train-\{clean-360,other-500\}}; and 
LS-100/TED3, 
an out-domain setting with unlabeled \texttt{\footnotesize train-ted3}.

\noindent\textbf{Training and decoding configurations:} All the experiments were conducted using ESPnet~\cite{watanabe2018espnet}. 
We used the Transformer architecture~\cite{vaswani2017attention} as E2E ASR model, 
which consisted of a stack of 12 self-attention layers with the same configurations as in~\cite{karita2019improving}\footnote{Note that the size of our model is relatively small compared to other state-of-the-art works on 
semi-supervised ASR~\cite{xu2020iterative, likhomanenko2020slimipl}. Future work will look to confirm the effectiveness of MPL on larger models.}. 
The base model was trained for 150 epochs using the Adam optimizer~\cite{kingma2015adam} with $\beta_1\!=\!0.9$, $\beta_2\!=\!0.98$, $\epsilon\!=\!10^{-9}$, and Noam learning rate scheduling~\cite{vaswani2017attention}.
We used 25,000 warmup steps and a learning rate factor of $5.0$.
The MPL training was iterated for 200 epochs and 
the online model was trained using the Adam optimizer with an initial learning rate of $10^{-3}$, $\beta_1\!=\!0.9$, $\beta_2\!=\!0.999$, and $\epsilon\!=\!10^{-8}$. 
For evaluation, a final model was obtained by averaging model parameters over 10 checkpoints with the best validation performance. 
For decoding with an LM, 
we trained one long short-term memory (LSTM) layer with 1024 units, 
using the \texttt{\footnotesize train-clean-100} transcriptions combined with the external text data provided by LibriSpeech~\cite{panayotov2015librispeech}. 
The LM was incorporated into decoding via shallow fusion 
with a beam-size of $20$ and an LM weight of $0.5$. 
For decoding without an LM, 
we performed the best path decoding of CTC~\cite{graves2006connectionist}. 
We used $w\!=\!0.5$ for all MPL experiments, 
leading to $\alpha\!=\!0.99977$ for LS-100/LS360, $\alpha\!=\!0.99989$ for LS-100/LS-860, and $\alpha\!=\!0.99983$ for LS-100/TED3.

\vspace{-.2cm}
\subsection{Main results}
\vspace{-.1cm}
Table \ref{tb:ls} shows results on the in-domain LS settings 
in terms of word error rate (WER) and WER recovery rate (WRR)~\cite{ma2008unsupervised}. 
Topline results are obtained via fully supervised training.
Looking at the LS-360 setting results (\texttt{A*}), 
both standard PL~\cite{kahn2020self} and proposed MPL led to a significant improvement over the base model (\texttt{L0}). 
MPL (\texttt{A3}) outperformed PL (\texttt{A1}) by dynamically updating pseudo-labels using the offline model, 
instead of fixing them to those generated by the base model~\cite{kahn2020self}. 
While PL could be improved by training on strong pseudo-labels generated via beam-search decoding with LM (\texttt{A2}),  
MPL achieved comparable performance without the help of LM (\texttt{A3}). 
Moreover, with LM, 
PL was prone to overfit to the LM training text %
as reported in~\cite{khurana2021unsupervised,likhomanenko2020slimipl}; 
in contrast, MPL successfully increased model generalization, 
achieving higher WRR of $81.2$\% on the ``other'' set. 
We also investigated an effective way for incorporating an LM into MPL, 
where PL is first applied to the base model with strong pseudo-labels (as in \texttt{A2}) and 
the pre-trained model is used for better initialization in MPL. 
With this pre-training method, 
MPL achieved the highest WRRs of $72.2$\%/$83.0$\% (\texttt{A4}), 
mitigating the over-fitting to the LM. 
In the LS-860 setting (\texttt{B*}), 
MPL again outperformed 
the baseline. %
While standard PL was not as effective as in the LS-360 setting in terms of WRR (\texttt{B1},\texttt{B2} vs.\ \texttt{A1},\texttt{A2}), 
MPL successfully recovered the same rate of errors (\texttt{B3},\texttt{B4} vs.\ \texttt{A3},\texttt{A4}),
demonstrating its scalability with respect to the amount of unlabeled data. 

Table \ref{tb:ted3} lists MPL results on the out-domain TED3 setting. 
Standard PL resulted in a modest improvement with WRR of at most $49.1$\% (\texttt{C1},\texttt{C2}). 
In contrast, 
the gain was more substantial for MPL with WRR of $80.8$\% (\texttt{C3},\texttt{C4}), 
indicating MPL is capable of efficiently adapting the base model to another domain. 
For PL, 
due to the domain mistmach, 
the advantage of utilizing the LM (trained on LS transcriptions) was smaller compared to the in-domain settings (\texttt{C2} vs.\ \texttt{A2},\texttt{B2}). 
MPL, however, succeeded in making use of the LM via the pre-training trick (\texttt{C4}). 

In all experiments, decoding with an LM further improved performance, with MPL again achieving high WRR.

\begin{table*}[t]
    \centering
    \caption{Word error rate (WER) [{\footnotesize $\%$}] and WER recovery rate (WRR) [{\footnotesize $\%{\uparrow}$}] on in-domain LibriSpeech (LS) settings. 
    The results are divided into two sections, depending on whether the LM with beam-search decoding was applied in the final evaluation or not.}\vspace{-.3cm}
    \centering
    \label{tb:ls}
         \sisetup{table-format=2.1,round-mode=places,round-precision=1,table-number-alignment = center,detect-weight=true,detect-inline-weight=math}
    \resizebox{.93\linewidth}{!}{
    \renewcommand{\arraystretch}{0.85}
    \begin{tabular}{lllSSSSSSSSSSSS}
    \toprule
    & & & \multicolumn{6}{c}{Decoding without LM} & \multicolumn{6}{c}{Decoding with LM} \\ 
    \cmidrule(l{0.3em}r{0.8em}){4-9} \cmidrule(l{0.3em}r{0.8em}){10-15}
    & & & \multicolumn{2}{c}{\textbf{Dev WER} [{\footnotesize $\%$}]} & \multicolumn{2}{c}{\textbf{Test WER} [{\footnotesize $\%$}]} & \multicolumn{2}{c}{\textbf{Test WRR} [{\footnotesize $\%{\uparrow}$}]} & \multicolumn{2}{c}{\textbf{Dev WER} [{\footnotesize $\%$}]} & \multicolumn{2}{c}{\textbf{Test WER} [{\footnotesize $\%$}]} & \multicolumn{2}{c}{\textbf{Test WRR} [{\footnotesize $\%{\uparrow}$}]} \\
    \cmidrule(l{0.3em}r{0.5em}){4-5} \cmidrule(l{0.3em}r{0.5em}){6-7} \cmidrule(l{0.3em}r{0.8em}){8-9} \cmidrule(l{0.3em}r{0.5em}){10-11} \cmidrule(l{0.3em}r{0.5em}){12-13} \cmidrule(l{0.3em}r{0.8em}){14-15}
    \textbf{Setting} & \textbf{Method} & \textbf{Init.} & {\footnotesize clean} & {\footnotesize other} & {\footnotesize clean} & {\footnotesize other} & {\footnotesize clean} & {\footnotesize other} & {\footnotesize clean} & {\footnotesize other} & {\footnotesize clean} & {\footnotesize other} & {\footnotesize clean} & {\footnotesize other} \\
    \midrule
    LS-100 & \texttt{L0}\hspace{2mm} baseline & -- & 13.2 & 31.1 & 13.5 & 32.4 & 0.0 & 0.0 & 8.8 & 23.6 & 9.1 & 24.5 & 0.0 & 0.0 \\
    \midrule
    \multirow{4}{*}[-8pt]{\shortstack[l]{LS-100\\\ / LS-360}} & \texttt{A1}\hspace{2mm} PL~\cite{kahn2020self} & \texttt{L0} & 10.5 & 25.2 & 10.7 & 25.8 & 45.3 & 54.5 & 7.3 & 18.7 & 7.6 & 19.4 & 39.2 & 52.4 \\
    & \texttt{A2}\hspace{2mm} PL~\cite{kahn2020self} & \texttt{L0} + LM & 9.1 & 23.4 & 9.4 & 23.9 & 66.5 & 70.0 & 7.0 & 18.2 & 7.1 & 18.6 & 51.3 & 61.4 \\
    & \texttt{A3}\hspace{2mm} MPL & \texttt{L0} & 9.4 & 22.4 & 9.6 & 22.6 & 63.6 & 81.2 & 7.0 & 17.2 & 7.2 & 17.5 & 48.7 & 72.3 \\
    & \texttt{A4}\hspace{2mm} MPL & \texttt{A2}$^\dagger$ & \bfseries 8.7 & \bfseries 22.0  & \bfseries 9.0 & \bfseries 22.4 & \bfseries 72.2 & \bfseries 83.0 & \bfseries 6.5 & \bfseries 16.9 & \bfseries 6.8 & \bfseries 17.1 & \bfseries 58.5 & \bfseries 76.4 \\
    \cdashlinelr{2-15}
    & \texttt{A5}\hspace{2mm} topline & \texttt{L0} & 6.7 & 20.2 & 7.3 & 20.3 & 100.0 & 100.0 & 4.8 & 15.0 & 5.1 & 14.9 & 100.0 & 100.0 \\ 
    \midrule
    \multirow{4}{*}[-8pt]{\shortstack[l]{LS-100\\\ / LS-860}} & \texttt{B1}\hspace{2mm} PL~\cite{kahn2020self} & \texttt{L0} & 10.4 & 24.3 & 10.7 & 25.0 & 37.0 & 41.4 & 7.3 & 18.0 & 7.4 & 18.4 & 34.4 & 43.5 \\
    & \texttt{B2}\hspace{2mm} PL~\cite{kahn2020self} & \texttt{L0} + LM & 8.7 & 21.4 & 8.9 & 21.9 & 59.4 & 58.5 & 6.6 & 16.6 & 6.9 & 17.0 & 45.6 & 52.8 \\
    & \texttt{B3}\hspace{2mm} MPL & \texttt{L0} & 9.0 & 18.0 & 9.2 & 18.1 & 56.0 & 80.0 & 6.9 & 13.7 & 7.0 & 14.1 & 44.2 & 73.7 \\
    & \texttt{B4}\hspace{2mm} MPL & \texttt{B2}$^\dagger$ & \bfseries 8.2 & \bfseries 17.5 & \bfseries 8.4 & \bfseries 17.6 & \bfseries 66.2 & \bfseries 83.0 & \bfseries 6.3 & \bfseries 13.5 & \bfseries 6.4 & \bfseries 13.7 & \bfseries 55.0 & \bfseries 76.8 \\
    \cdashlinelr{2-15}
    & \texttt{B5}\hspace{2mm} topline & \texttt{L0} & 5.7 & 14.4 & 5.8 & 14.6 & 100.0 & 100.0 & 4.1 & 10.5 & 4.3 & 10.4 & 100.0 & 100.0 \\
    \bottomrule
    \end{tabular}}\\
    {\footnotesize $^\dagger$We used the model from 100 epochs and applied MPL for 100 epochs so that the total number of training epochs matches that of the other methods.}
    \vspace{-.3cm}
\end{table*}

\begin{table*}[t]
    \centering
    \caption{WER [{\footnotesize $\%$}] and WRR [{\footnotesize $\%{\uparrow}$}] on out-domain TEDLIUM3 (TED3) setting.}\vspace{-.3cm}
    \centering
    \label{tb:ted3}
      \sisetup{table-format=2.1,round-mode=places,round-precision=1,table-number-alignment = center,detect-weight=true,detect-inline-weight=math}
    \resizebox{.9\linewidth}{!}{
    \renewcommand{\arraystretch}{0.85}
    \begin{tabular}{lllS@{\hspace{1\tabcolsep}}S@{\hspace{1\tabcolsep}}S@{\hspace{1\tabcolsep}}S@{\hspace{1\tabcolsep}}S@{\hspace{1\tabcolsep}}S}
    \toprule
    & & & \multicolumn{3}{c}{Decoding without LM} & \multicolumn{3}{c}{Decoding with LM} \\ 
    \cmidrule(lr{0.5em}){4-6} \cmidrule(l{-0.1em}r{0.5em}){7-9}
    \textbf{Setting} & \textbf{Method} & \textbf{Init.} & \textbf{Dev WER} [{\footnotesize $\%$}] & \textbf{Test WER} [{\footnotesize $\%$}] & \textbf{Test WRR} [{\footnotesize $\%{\uparrow}$}] & \textbf{Dev WER} [{\footnotesize $\%$}] & \textbf{Test WER} [{\footnotesize $\%$}] & \textbf{Test WRR} [{\footnotesize $\%{\uparrow}$}] \\
    \midrule
    LS-100 & \texttt{L0}\hspace{2mm} baseline & -- & 32.5 & 33.2 & 0.0 & 26.8 & 26.8 & 0.0 \\
    \midrule
    \multirow{4}{*}[-8pt]{\shortstack[l]{LS-100\\\ / TED3}} & \texttt{C1}\hspace{2mm} PL~\cite{kahn2020self} & \texttt{L0} & 26.8 & 26.0 & 35.6 & 22.1 & 20.9 & 34.6 \\
    & \texttt{C2}\hspace{2mm} PL~\cite{kahn2020self} & \texttt{L0} + LM & 24.4 & 23.2 & 49.1 & 21.0 & 20.1 & 39.2 \\
    & \texttt{C3}\hspace{2mm} MPL & \texttt{L0} & 18.8 & 17.6 & 76.3 & 16.8 & 15.5 & 65.9 \\
    & \texttt{C4}\hspace{2mm} MPL & \texttt{C2} & \bfseries 18.2 & \bfseries 16.7 & \bfseries 80.8 & \bfseries 16.2 & \bfseries 14.9 & \bfseries 69.6 \\
    \cdashlinelr{2-9}
    & \texttt{C5}\hspace{2mm} topline & \texttt{L0} & 12.7 & 12.8 & 100.0 & 10.0 & 9.6 & 100.0 \\
    \bottomrule
    \end{tabular}}\vspace{-.45cm}
\end{table*}

\vspace{-.2cm}
\subsection{Online model vs. offline model}
\vspace{-.1cm}
\label{ssec:on_vs_off}
Contrary to previous work~\cite{tarvainen2017mean}, 
we adopted the online model for final evaluation. 
When we did not use the checkpoint averaging technique~\cite{vaswani2017attention}, 
the offline model gave better performance than the online model (e.g., $9.4$\%/$22.7$\% vs.\ $9.8$\%/$24.0$\% on dev.\ sets in the LS-360 setting). 
Since the offline model is an average of the online models over the training (cf.~\eqref{eq:ema}), 
the offline model naturally benefited from the model ensembling. 
However, 
with checkpoint averaging, 
both of the models were improved and 
the performance gap was reduced to almost none (e.g., $9.4$\%/$22.5$\% vs.\ $9.4$\%/$22.4$\% on dev.\ sets in the LS-360 setting). 
We used the slightly better online model for evaluation.

\vspace{-.2cm}
\subsection{Importance of data augmentation}
\vspace{-.1cm}
\label{ssec:exp_data_augmentation}
When applying MPL to the base model (\texttt{L0} in Table \ref{tb:ls}) without SpecAugment, 
WRRs dropped to $39.7$\%/$48.1$\% compared to the results with SpecAugment (\texttt{A3}). 
Note that, for the experiment without SpecAugment, 
the WRRs were calculated based on a topline model trained without the augmentation. 
Without SpecAugment, 
we observed that the MPL training converged earlier and 
MPL was less effective.

\vspace{-.2cm}
\subsection{Effectiveness of $w$ for tuning the momentum update}
\vspace{-.1cm}
\label{ssec:exp_momentum_update}
\begin{figure}[t]
    \centering
    \footnotesize
    \begin{tabular}{c}
        \hspace{-0.35cm}
        \begin{minipage}{0.5\columnwidth}
            \centering
            \subcaption*{\scriptsize \hspace{0.65cm}{\rm (a)} \textbf{LS-100 / LS-360}}
            \vspace{-0.3cm}
            \includegraphics[width=\linewidth]{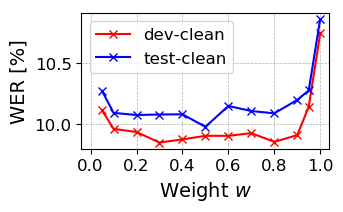}
            \vspace{-0.7cm}
        \end{minipage}
        \begin{minipage}{0.5\columnwidth}
            \centering
            \subcaption*{\scriptsize \hspace{0.65cm}{\rm (b)} \textbf{LS-100 / LS-860}}
            \vspace{-0.3cm}
            \includegraphics[width=\linewidth]{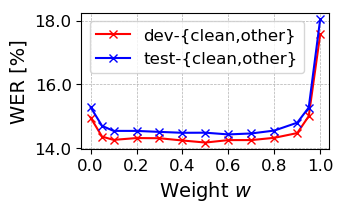}
            \vspace{-0.7cm}
        \end{minipage}\\\\
        \hspace{-0.35cm}
        \begin{minipage}{0.5\columnwidth}
            \centering
            \subcaption*{\scriptsize \hspace{0.6cm}{\rm (c)} \textbf{LS-100 / TED3}}
            \vspace{-0.3cm}
            \includegraphics[width=\linewidth]{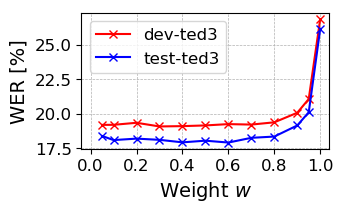}
        \end{minipage}
        \begin{minipage}{0.5\columnwidth}
            \centering
            \subcaption*{\scriptsize \hspace{0.65cm}{\rm (d)} \textbf{LS-100 / TED3 (unsup)}}
            \vspace{-0.3cm}
            \includegraphics[width=\linewidth]{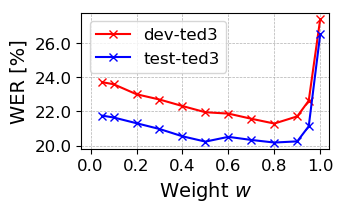}
        \end{minipage}
    \end{tabular}\vspace{-.3cm}
    \caption{Influence of momentum update weight $w$ on WER.}
    \label{fig:w-wer}%
\end{figure}
Figure~\ref{fig:w-wer} shows the model performance depending on the weight $w$ used to calculate $\alpha$ in the momentum update \eqref{eq:ema}. 
Here, 
we observed a similar trend among the curves in different semi-supervised settings (Figs.\ \ref{fig:w-wer}(a), \ref{fig:w-wer}(b), and \ref{fig:w-wer}(c)). 
The performance degraded as $w$ was set closer to $0.0$. 
When $w=0.0$, which is a similar approach to~\cite{chen2020semi}, 
the training was likely to be unstable and failed under the conditions in Figs.\ \ref{fig:w-wer}(a) and \ref{fig:w-wer}(c), 
indicating the importance of retaining parameters from the base model. 
However, 
depending too much on the base model (i.e., setting $w$ closer to $1.0$) also worsened performance, since larger $w$ slows down the offline model's progress, 
causing MPL to become more like standard PL~\cite{kahn2020self}. 
Fig.~\ref{fig:w-wer}(d) shows results under an extreme condition, 
where not only a domain mismatch exists between the base model and unlabeled data 
but labeled data is not used during the semi-supervised process (i.e., training with \eqref{eq:l_unsup} only). 
The performance was more sensitive to the change in $w$ than in the other settings, but the overall trend was similar. 

Overall, the proposed tuning method effectively controlled the momentum update in all settings. 
It provides a more intuitive guide for tuning $\alpha$, taking the amount of data into account. 
Based on the validation results mainly on the LS-860 and TED3 settings, 
we set $w=0.5$ for all the semi-supervised conditions.

\vspace{-.2cm}
\subsection{Comparison with iterative pseudo-labeling}
\vspace{-.1cm}
\begin{table}[t]
    \centering
    \caption{Comparison of test WER [{\footnotesize $\%$}] between iterative PL and MPL in each setting. PL\# denotes PL at the \#-th iteration. } \vspace{-.3cm}
    \centering
    \label{tb:ipl}
     \sisetup{table-format=2.1,round-mode=places,round-precision=1,table-number-alignment = center,detect-weight=true,detect-inline-weight=math}
    \resizebox{.97\linewidth}{!}{
    \renewcommand{\arraystretch}{0.85}
    \begin{tabular}{lcSSSSS}
    \toprule
    \textbf{Setting} & \textbf{Test data}& \textbf{PL1} & \textbf{PL2} & \textbf{PL3} & \textbf{PL4} & \textbf{MPL} \\
    \cmidrule(lr{0.5em}){1-2} \cmidrule(lr{0.5em}){3-6} \cmidrule(lr{0.5em}){7-0}
    \multirow{2}{*}[0pt]{LS-100 / LS-360} & test-clean & 11.3 & 10.5 & 10.2 & 9.9 & \bfseries 9.6 \\
    & test-other & 27.3 & 25.5 & 24.6 & 24.2 & \bfseries 22.6 \\
    \midrule
    \multirow{2}{*}[0pt]{LS-100 / LS-860} & test-clean & 11.1 & 10.3 & 9.7 & 9.4 & \bfseries 9.2 \\
    & test-other & 25.6 & 23.2 & 21.9 & 20.7 & \bfseries 18.1 \\
    \midrule
    LS-100 / TED3 & test-ted3\phantom{r} & 26.3 & 23.5 & 22.2 & 21.3 & \bfseries 17.6 \\
    \bottomrule
    \end{tabular}}
\end{table}
In Table \ref{tb:ipl}, we list results comparing MPL with a simple iterative PL (IPL) method similar to~\cite{xu2020iterative, likhomanenko2020slimipl}. 
In IPL, 
the base model was continuously trained on pseudo-labels 
which were updated at intervals of 50 epochs. 
For a fair comparison, 
the update was performed four times to match the total number of epochs to that of MPL (i.e., 200 epochs). 
The results show the effectiveness of IPL, 
as performance gradually improved with the iterations. 
However, MPL still 
outperformed the final round of PL, 
indicating MPL is more effective at training with unlabeled data. 

\vspace{-.2cm}
\section{Conclusions}
\vspace{-.1cm}
This paper proposed MPL, 
momentum pseudo-labeling for semi-supervised ASR. 
Experimental results on various semi-supervised settings demonstrated its effectiveness,
achieving clear improvements over standard pseudo-labeling and iterative pseudo-labeling. %
Moreover, MPL was shown to be %
effective independently of the amount of unlabeled data or domain mismatch. 
Future work includes applying filtering techniques ~\cite{khurana2021unsupervised} and introducing multiple hypotheses~\cite{moritz2021semi} in the MPL process.

\bibliographystyle{IEEEtran}
\bibliography{refs}

\end{document}